\documentclass[reqno,11pt]{amsart}
\usepackage[utf8]{inputenc}
\usepackage{enumitem}
\usepackage{esint}
\usepackage{graphicx}
\usepackage{amscd}
\usepackage{slashed}
\usepackage{amssymb}
\usepackage{mathtools} 
\usepackage[mathscr]{eucal}
\numberwithin{equation}{section}
\allowdisplaybreaks[1]

\title[A Canonical Complex Structure for Scalar Fields]
{A Canonical Complex Structure and the Bosonic Signature Operator
for Scalar Fields \\ in Globally Hyperbolic Spacetimes}

\author[F.\ Finster]{Felix Finster}
\address{Fakult\"at f\"ur Mathematik \\ Universit\"at Regensburg \\ D-93040 Regensburg \\ Germany}
\email{finster@ur.de}

\author[A.\ Much]{Albert Much \\ \\ December 2021}
\address{Institut f\"ur Theoretische Physik\\ Universit\"at Leipzig\\ D-04103 Leipzig \\ Germany}
\email{much@itp.uni-leipzig.de}

\newtheorem{Def}{Definition}[section]
\newtheorem{Thm}[Def]{Theorem}
\newtheorem{Prp}[Def]{Proposition}
\newtheorem{Lemma}[Def]{Lemma}

\newcommand{\Thanks}{\vspace*{.5em} \noindent \thanks}
\newcommand{\beq}{\begin{equation}}
\newcommand{\eeq}{\end{equation}}
\newcommand{\Proof}{\begin{proof}}
\newcommand{\QED}{\end{proof} \noindent}

\newcommand{\la}{\langle}
\newcommand{\ra}{\rangle}

\newcommand{\C}{\mathbb{C}}
\newcommand{\R}{\mathbb{R}}
\newcommand{\1}{\mbox{\rm 1 \hspace{-1.05 em} 1}}

\renewcommand{\H}{\mathscr{H}}

\newcommand{\bep}{\begin{pmatrix}}
\newcommand{\enp}{\end{pmatrix}}

\newcommand{\Lin}{\text{\rm{L}}}
\newcommand{\loc}{\text{\rm{loc}}}
\newcommand{\Cisc}{C^\infty_{\text{\rm{sc}}}}
\newcommand{\Cisco}{C^\infty_{\text{\rm{sc}},0}}
\newcommand{\bitem}{\begin{itemize}[leftmargin=2em]}
\newcommand{\eitem}{\end{itemize}}
\newcommand{\x}{\mathbf{x}}

\newcommand{\hol}{\text{\rm{hol}}}
\newcommand{\ah}{\text{\rm{ah}}}

\DeclareMathOperator{\im}{Im}

\DeclareMathOperator{\supp}{supp}

\newcommand{\p}{\mathfrak{p}}
\newcommand{\Sig}{\mathscr{S}}

\newcommand{\scrM}{\mycal M}
\newcommand{\scrN}{\mycal N}
\newcommand{\Sol}{\mathrm{Sol}}

\DeclareFontFamily{OT1}{rsfso}{}
\DeclareFontShape{OT1}{rsfso}{m}{n}{ <-7> rsfso5 <7-10> rsfso7 <10-> rsfso10}{}
\DeclareMathAlphabet{\mycal}{OT1}{rsfso}{m}{n}

\begin{document}

\maketitle
\begin{abstract}
The bosonic signature operator is defined for Klein-Gordon fields and massless scalar fields on globally
hyperbolic Lorentzian manifolds of infinite lifetime.
The construction is based on an analysis of
families of solutions of the Klein-Gordon equation with a varying mass parameter.
It makes use of the so-called bosonic mass oscillation property which states that
integrating over the mass parameter generates decay of the field at infinity.
We derive a canonical decomposition of the solution space of the Klein-Gordon equation
into two subspaces, independent of observers or the choice of
coordinates. This decomposition endows the solution space with a canonical complex structure.
It also gives rise to a distinguished quasi-free state.
Taking a suitable limit where the mass tends to zero, we obtain
corresponding results for massless fields.
Our constructions and results are illustrated in the examples of Minkowski space and
ultrastatic spacetimes.
\end{abstract}

\tableofcontents

\section{Introduction}
Quantum field theory in globally hyperbolic spacetimes plays an important role in
describing physical systems in the presence of gravitation fields, with applications
to quantum effects near black holes and to cosmology.
Furthermore, it can be regarded as an intermediate step towards quantum gravity,
i.e.\ a unified description of spacetime and matter in the quantum realm.
In the framework of algebraic quantum field theory, the quantization scheme consists of two steps (see for example~\cite{benini+dappiaggi+hack,brunettibook, gerard-intro}).
First, one associates to a physical system a unital $*$-algebra $\mathcal{A}$, whose elements are interpreted as observables and which encodes structural properties such as causality and the canonical commutation or anti-commutation relations. In the second step, one chooses a {\em{state}},
defined as a positive linear functional~$\omega:\mathcal{A}\to\mathbb{C}$. Out of the pair $(\mathcal{A},\omega)$ one recovers the standard probabilistic interpretation of quantum theory via the GNS theorem, which gives a representation of the algebra by operators on a Hilbert space.

From the variety of states, one wants to select those states which are physically sensible.
These states are characterized mathematically by the so-called {\em{Hadamard condition}}, a constraint on the wavefront set of the underlying two point distribution (see \cite{gerard-intro,brunettibook} for recent surveys on this topic). While the existence of Hadamard states can be shown abstractly with glueing constructions
(see~\cite{fulling+sweeny+wald, fulling+narcowich+wald}),
these methods are not explicit. Therefore, the construction of Hadamard states
is an ongoing subject of intensive research.
In recent years, two methods for constructing Hadamard states have been developed:
the pseudo-differential operator approach~\cite{gerard2014construction,
gerard+wrochna2, gerard+wrochna3} and the holographic method~\cite{dappiaggi+moretti, dappiaggi+hack+pinamonti, dappiaggi+siemssen, benini+dappiaggi+murro}.
The first method gives a whole class of states.
The second method does give a distinguished state, but it has the shortcoming that it relies on
conformal methods which only apply to massless fields.

For Dirac fields, a distinguished state which does not depend on observers or the choice of
coordinates is the so-called {\em{fermionic projector state}}. Its construction uses the splitting
of the Dirac solution space corresponding to the sign of the spectrum of the so-called
{\em{fermionic signature operator}}~$\Sig_m$.
This construction goes back to the perturbative treatment of the Dirac sea in~\cite{sea}.
The fermionic signature operator was first introduced in spacetimes of finite lifetime in~\cite{finite}.
The construction has been extended in~\cite{infinite} to spacetimes of infinite lifetime, making use of the
so-called {\em{mass oscillation property}}. This construction has been
proven to be useful for the construction of Hadamard states~\cite{hadamard, planewave, desitter}.
Moreover, the fermionic signature operator respects the spacetime symmetries~\cite{sigsymm}
and is a suitable starting point for spectral geometry with Lorentzian signature~\cite{drum}.
Generalizations and related constructions can be found in~\cite{rindler, sigbh, drago+murro}.

It is the goal of the present paper to adapt the construction of the fermionic signature operator
to {\em{bosonic fields}}. For simplicity, we restrict attention to scalar fields,
i.e.\ to the Klein-Gordon equation 
\[ \big( \Box_g+m^2+\xi R_g \big) \phi=0\:, \]
and the massless wave equation
$$ \Box_g \phi=0.$$ 
with real parameters~$m$ and~$\xi$.
In this context, similar constructions in spacetimes of finite lifetime give rise to so-called
SJ-states (see~\cite{sorkin, sorkin2, johnston} and~\cite{brum-fredenhagen, FV}).
In spacetimes of infinite lifetime, the only related constructions concern the
analysis of Feynman propagators in~\cite{derezinski-siemsen2}.
In the present paper, we introduce the mass oscillation property for the Klein-Gordon field
and construct the corresponding {\em{bosonic signature operator}}.
As an application, we show that the bosonic signature operator   endows the solution
space of the Klein-Gordon equation with a {\em{canonical complex structure}}.
It also gives rise to a distinguished quasi-free state, the {\em{bosonic projector state}}.
Moreover, we show that, under suitable assumptions, one can take the limit~$m \searrow 0$
to extend the results to {\em{massless fields}}.
We remark that our methods and results generalize in a straightforward way to
other bosonic equations, in particular to the wave equation in the presence of gauge fields.

The main obstacle for adapting the constructions from the Dirac field to the Klein-Gordon field
is that, on the space of solutions of the Klein-Gordon equation, there is no canonical scalar product,
but instead the symplectic form~$\sigma_m$ defined by
\beq \label{sigmaintro}
\sigma_m(\phi_m, \tilde{\phi}_m) := \int_\scrN \Big( 
\overline{\partial_j \phi_m}\: \tilde{\phi}_m - \overline{\phi_m}\: \partial_j \tilde{\phi}_m \Big) \: \nu^j \:d\mu_\scrN(x) \:,
\eeq
where~$\scrN$ is an arbitrary Cauchy surface. As a consequence, the solution space does not carry a natural
topology. In order to get a canonical scalar product and topology, we want to make use of the $L^2$-scalar product
in spacetime after integrating over the mass parameter.
Our goal is to derive a {\em{mass decomposition}} of this scalar product of the form
\beq \label{massintro}
\bigg\la \int_I \phi_m \,m\,dm,  \int_I \tilde{\phi}_{m'} \,m'\,dm' \bigg\ra_{L^2(\scrM, d\mu_\scrM)} 
= \int_I \la \phi_m | \tilde{\phi}_m \ra_m \:m\,dm \:,
\eeq
where~$(\phi_m)_{m \in I}$ and~$(\tilde{\phi}_m)_{m \in I}$ are families of solutions
of the Klein-Gordon equation for a mass parameter~$m \in I$ with~$0 \not \in \overline{I}$
with suitable regularity properties (for details see Section~\ref{secMOP}).
We remark that the factors~$m$ and~$m'$ in the above integrals can be understood from the fact that
the Klein-Gordon equation only involves the mass squared. Therefore, it is preferable to
integrate over the square of the mass~$a:=m^2$, which in view of the transformation law~$da = 2m \,dm$
of the integration measures gives rise to the factor of~$m$.

Once the decomposition~\eqref{massintro}
has been obtained, the scalar product~$\la .|. \ra_m$ in the integrand on the right hand side
endows the solution space of the Klein-Gordon equation of mass~$m$
with the desired canonical Hilbert space structure.
The main technical difficulty in implementing this method
is that a Hilbert space structure on the solution space is needed
right from the beginning in order to derive the above mass decomposition.
In order to overcome this difficulty, we introduce an {\em{auxiliary scalar product}} of the form
\[ \la \phi_m \,|\, \tilde{\phi}_m \ra_{A_m} := i \sigma_m \big( \phi_m, A_m \tilde{\phi}_m \big) \:, \]
where~$(A_m)_{m \in I}$ is a suitable family of operators on the symplectic space
(for details see again Section~\ref{secMOP}).
Then we construct the mass decomposition~\ref{massintro} and show that
it is indeed independent of the choice of the auxiliary scalar product.
Moreover, we prove that the bosonic signature operator is uniquely defined as
a bounded linear operator on the Hilbert space~$(\H_m, \la .|. \ra_m)$.

In order to illustrate our methods and results, we consider two explicit examples:
The Klein-Gordon equation in Minkowski space and the Klein-Gordon equation in an ultrastatic
spacetimes. By explicit computation, we obtain in both cases that the bosonic signature operator
reproduces the splitting of the solution space into solutions of positive and negative frequency.
In this way, we obtain consistency with the usual construction of states in these spacetimes.
The main benefit of using the bosonic signature operator is that the construction applies
also in situations when spacetime is not ultrastatic, giving rise to a distinguished quantum state
which is defined using the global geometry of spacetime.
Other future applications include the study of the entanglement entropy
in the absence of Killing fields.

The paper is organized as follows. In Section~\ref{secprelim} we provide the necessary background
on globally hyperbolic spacetimes and the Klein-Gordon equation.
In Section~\ref{secMOP} the bosonic mass oscillation property is defined and the corresponding
mass decomposition is derived.
In Section~\ref{secsig}, the resulting bosonic signature operator is introduced,
and it is shown that it gives rise to a canonical complex structure on the solution space
of the Klein-Gordon equation as well as to the bosonic projector state.
In Section~\ref{secmassless}, it is explained how the massless case can be treated
by taking a suitable limit~$m \searrow 0$.
In Section~\ref{secmink} it is shown that
in flat Minkowski space, the bosonic signature operator has eigenvalues~$\pm 1$, and that the
corresponding eigenspaces are the subspaces of positive and negative frequency, respectively.
In Section~\ref{secultra} it is shown that this holds true in general ultrastatic spacetimes.

\section{Preliminaries} \label{secprelim}
Let~$(\scrM, g)$ be a smooth, globally hyperbolic Lorentzian manifold of dimension~$k \geq 2$.
For the signature of the metric we use the convention~$(+ ,-, \ldots, -)$.
As proven in~\cite{bernal+sanchez}, $\scrM$ admits a smooth foliation~$(\scrN_t)_{t \in \R}$
by Cauchy hypersurfaces. Thus~$\scrM$ is diffeomorphic to the product of~$\R$ with a
$(k-1)$-dimensional manifold.
We denote the wave operator on~$(\scrM, g)$ by~$\Box_g$,
\[ \Box_g \phi := \frac{1}{\sqrt{-\det g}} \: \frac{\partial}{\partial x^i} \Big( \sqrt{-\det g} \:g^{ij} \partial_j \phi \Big) \:, \]
where $\phi \in C^\infty(\scrM) = C^\infty(\scrM, \C)$ is a smooth, complex scalar field. The {\em{Klein-Gordon operator}}~$P_m$ is defined by
\begin{align}\label{kge}
P_m\phi:= \big( \Box_g+m^2+\xi R_g \big) \phi=0\:,
\end{align}
where~$m$ is the rest mass and~$\xi \in \R$ is a dimensionless coupling constant
to scalar curvature~$R_g$. We denote the solutions of the Klein-Gordon equation by~$\Sol_m(\scrM)$.
For clarity we remark that we always consider {\em{complex-valued}} fields.

The Cauchy problem for the Klein-Gordon equation is well-posed. 
This can be seen either by considering energy estimates for symmetric hyperbolic systems
(see for example~\cite{john} or~\cite[Chapter~13]{intro}) or alternatively by constructing the Green's kernel (see for
example~\cite{baer+ginoux}). These methods also show that the Klein-Gordon equation is causal,
meaning that the solution of the Cauchy problem only depends on the initial data in the causal
past or future. In particular, if the initial data on~$\scrN_t$ has compact support,
the solution~$\phi$ also has compact support on any other Cauchy hypersurface.
This leads us to consider solutions~$\phi_m$ in the class~$\Cisc(\scrM) \cap \Sol_m(\scrM)$ of smooth functions with
spatially compact support. On solutions in this class,
one introduces the symplectic form~$\sigma_\scrN$ by
\beq
\begin{split}
&\sigma_\scrN : \Cisc(\scrM) \cap \Sol_m(\scrM) \times \Cisc(\scrM) \cap \Sol_m(\scrM) \rightarrow \C \:, \\
&\sigma_\scrN(\phi_m, \tilde{\phi}_m) := \int_\scrN \Big( 
\overline{\partial_j \phi_m}\: \tilde{\phi}_m - \overline{\phi_m}\: \partial_j \tilde{\phi}_m \Big) \: \nu^j \:d\mu_\scrN(x) \:,
\end{split} \label{sigma}
\eeq
where~$\nu$ denotes the future-directed normal. One immediately verifies that
the symplectic form is sesquilinear and skew-symmetric, i.e.\
\[ \sigma_\scrN(\phi_m, \tilde{\phi}_m) = - \overline{\sigma_\scrN(\tilde{\phi}_m,\phi_m )} \:. \]
Moreover, it does not depend on the choice of the
Cauchy surface~$\scrN$. In order to see this, we let~$\scrN'$ be another Cauchy surface and~$\Omega$
the spacetime region enclosed by~$\scrN$ and~$\scrN'$. We then obtain
\beq \begin{split} \label{divfree}
\nabla_j \Big( 
\overline{\partial^j \phi_m}\: \tilde{\phi}_m - \overline{\phi_m}\: \partial^j \tilde{\phi}_m \Big)
&=  \big( \overline{\Box_g \phi_m} \big)\: \tilde{\phi}_m - \overline{\phi_m}\: \big(\Box_g \tilde{\phi}_m\big) 
 \\
&=\big( \overline{P_m \phi_m} \big)\: \tilde{\phi}_m - \overline{\phi_m}\: \big(P_m \tilde{\phi}_m\big) = 0\:,
\end{split}
\eeq
where in the last step we used that~$\Box$ and~$P_m$ differ by a real-valued multiplication operator.
Integrating over~$\Omega$ and applying the Gau{\ss} divergence
theorem, we find that~$\sigma_\scrN(\psi_m, \phi_m) = \sigma_{\scrN'}(\psi_m, \phi_m)$.
In view of the independence of the choice of the Cauchy surface, we also denote
the symplectic form~\eqref{sigma} by~$\sigma_m(.,.)$.

The {\em{retarded}} and {\em{advanced Green's operators}}~$S_m^\wedge$ and~$S_m^\vee$ are formal adjoints of each other (with respect to the $L^2$-scalar product on~$\scrM$)
and linear mappings (for details see for example~\cite{kay-wald, baer+ginoux})
\[ S_m^\wedge, S_m^\vee \::\: C^\infty_0(\scrM) \rightarrow \Cisc(\scrM)\:, \]
where $C^\infty_0(\scrM)$ denotes the vector space of smooth functions with compact support. 
They satisfy the defining equation of the Green's operator
\[ 
P_m \big( S_m^{\wedge, \vee} f \big) = f \:. \]
Moreover, they are uniquely determined by the condition that the support of~$S_m^\wedge \phi$
(or~$S_m^\vee \phi$) lies in the causal future (respectively the causal past) of~$\supp \phi$.

The {\em{causal fundamental solution}} is defined as the difference of the advanced and retarded
Green's operators
\beq \label{Kdef}
G_m := S_m^\wedge - S_m^\vee \::\: C^\infty_0(\scrM) \rightarrow \Cisc(\scrM) \cap \text{Sol}_m(\scrM) \:;
\eeq
note that it maps to solutions of the Klein-Gordon equation.
The fundamental solution gives rise to a sesquilinear form, which we again denote by~$G_m$,
\beq \label{Gmform}
G_m \,:\, C^\infty_0(\scrM) \times C^\infty_0(\scrM) \rightarrow \C \:,\qquad
G_m(\phi, \psi) := \int_\scrM \overline{\phi(x)}\: (G_m \psi)(x)\: d\mu_\scrM(x) \:.
\eeq
In the next lemma, this sesquilinear is related to the symplectic form.
\begin{Lemma} For any~$f,g \in C^\infty_0(\scrM)$
\[ G_m(f,g) = \sigma_m\big( G_m f, G_m g \big) \:. \]
\end{Lemma}
\Proof Knowing that the definition of the symplectic form~\eqref{sigma} does not depend on the choice
of the Cauchy surface, it suffices to compute it for a Cauchy surface~$\scrN$ 
which lies in the future of the supports of~$f$ and~$g$. Then, using~\eqref{Kdef} and the
causal properties of the Green's operators, we obtain
\[ \sigma_m\big( G_m f, G_m g \big) = \sigma_\scrN\big( S^\wedge_m f, S^\wedge_m g \big)
= \int_\scrN \Big( \overline{\partial_j S^\wedge_m f}\: S^\wedge_m g
- \overline{S^\wedge_m f}\: \partial_j S^\wedge_m g \Big) \: \nu^j \:d\mu_\scrN(x) \:. \]
Choosing~$\Omega$ as the past of~$\scrN$, we can apply the Gauss divergence theorem 
and proceed as in~\eqref{divfree} to obtain
\begin{align*}
\sigma_m\big( G_m f, G_m g \big) &= \int_\Omega \nabla_j \Big( \overline{\partial^j S^\wedge_m f}\: S^\wedge_m g
- \overline{S^\wedge_m f}\: \partial^j S^\wedge_m g \Big)\: d\mu_\scrM \\
&= \int_\Omega \Big( \overline{P_m S^\wedge_m f}\: S^\wedge_m g - \overline{S^\wedge_m f}\: P_m S^\wedge_m g \Big)\: d\mu_\scrM \\
&= \int_\Omega \Big( \overline{f}\: S^\wedge_m g - \overline{S^\wedge_m f}\: g \Big)\: d\mu_\scrM
= \int_\Omega \Big( \overline{f}\: (S_m^\wedge - S^\vee_m) g \Big)\: d\mu_\scrM = G(f,g) \:.
\end{align*}
This gives the result.
\QED
This lemma shows in particular that~$G_m$ is skew-symmetric, i.e.
\[ \overline{G_m(f,g)} = -G_m(g,f)\:. \]

\section{The Bosonic Mass Oscillation Property} \label{secMOP}
We consider the mass parameter in a bounded open interval, $m \in I := (m_L, m_R)$
with~$0 \not\in \overline{I}$.
We consider families of solutions~$(\phi_m)_{m \in I}$ which are smooth and have spatially
compact support, i.e.\ $\phi_m \in \Cisc(\scrM) \cap \Sol_m(\scrM)$ for all~$m \in I$.
Moreover, we assume that the family depends smoothly on the mass parameter
and vanishes identically for~$m$ outside a compact subset of~$(m_L, m_R)$.
Writing the mass parameter as an additional argument (i.e.\ $\phi(x,m) = \phi_m(x)$),
we summarize these properties by writing
\[ 
\phi \in \Cisco(\scrM \times I) \:, \]
where~$\Cisco(\scrM \times I)$ denotes the smooth wave functions with spatially compact support which
are also compactly supported in~$I$.
We denote the operator of multiplication with~$m$ by~$T$,
\[ T \::\: \Cisco(\scrM \times I) \rightarrow \Cisco(\scrM \times I) \:,\qquad (T \phi)_m = m \,\phi_m \:. \]
Moreover, we denote the operation of integrating over~$m$ by~$\p$,
\beq \label{pdef}
(\p \psi)(x) = \int_I \psi_m(x)\: m\,dm \:,
\eeq
where, as explained after~\eqref{massintro}, we work with the integration measure~$m\,dm$
(and~$dm$ is the Lebesgue measure).
We assume that~$\p$ maps to the square-integrable functions in spacetime, i.e.\
\[ \p \::\: \Cisco(\scrM \times I) \rightarrow \Cisc(\scrM) \cap L^2(\scrM, d\mu_\scrM) \]
(this assumption will be justified by the bosonic mass oscillation property to be introduced below).
This makes it possible to introduce on~$\Cisco(\scrM \times I)$ a positive semi-definite
sesquilinear form~$\la .|. \ra_\H$ by
\[ \la .|. \ra_\H : \Cisco(\scrM \times I) \times \Cisco(\scrM \times I) \rightarrow \C \:, \qquad
\la \phi \,|\, \tilde{\phi} \ra_\H := \la \p \phi \,|\, \p \tilde{\phi} \ra_{L^2(\scrM, d\mu_\scrM)} \:. \]
Dividing out the null space and forming the completion gives a Hilbert space~$(\H, \la .|. \ra_\H)$.
We denote the corresponding norm by~$\| . \|_\H$. We also refer to the scalar product~$\la .|. \ra_\H$
as the {\em{spacetime scalar product}}.

We now introduce an {\em{auxiliary scalar product}} on the solutions of the Klein-Gordon equation.
\begin{Def} \label{def31}
For every~$m \in I$ we let~$A_m$ be a linear operator on~$\Cisc(\scrM) \cap \Sol_m(\scrM)$.
The family~$(A_m)_{m \in I}$ is referred to as a family of {\bf{auxiliary operators}} if the following
conditions holds:
\bitem
\item[{\rm{(i)}}] The sesquilinear form~$\la .|. \ra_{A_m}$ defined by
\beq \label{Amdef}
\la \phi_m \,|\, \tilde{\phi}_m \ra_{A_m} := i \sigma_m \big( \phi_m, A_m \tilde{\phi}_m \big)
\eeq
is positive semi-definite.
\item[{\rm{(ii)}}] The operators~$A_m$ are uniformly bounded in the sense that there
is a constant~$c>0$ such that
\beq \label{unibound}
\big| \sigma_m \big( A_m \phi_m, A_m \phi_m \big) \big| \leq c\: i \sigma(\phi_m, A_m \phi_m) 
\qquad \text{for all~$\phi \in \Cisco(\scrM)$} \:.
\eeq
\item[{\rm{(iii)}}] For any~$\phi, \tilde{\phi} \in \Cisco(\scrM \times I)$, the function
\beq \label{contin}
\sigma_m(\phi_m, A_m \tilde{\phi}_m) \qquad \text{is continuous in~$m \in I$}\:.
\eeq
\eitem
\end{Def} \noindent
We remark that a sesquilinear, positive semi-definite form is Hermitian in the sense that
\[ \overline{\la \psi | \phi \ra_{A_m}} = \la \phi | \psi \ra_{A_m} \:. \]
Thus, dividing out the null space, the sesquilinear form~$\la .|.\ra_{A_m}$ defines a scalar product.
Forming the completion
gives the Hilbert spaces~$(\H_{A_m}, \la .|. \ra_{A_m})$. We denote the corresponding norm by~$\|.\|_{A_m}$.

Let us explain the above assumptions.
The positivity of~\eqref{Amdef} implies that the operator~$A_m$ is {\em{symmetric}} on~$\H_{A_m}$, as the
following consideration shows. For any~$\phi_m \in \Cisc(\scrM) \cap \Sol_m(\scrM)$,
\beq \label{topolarize}
\R \ni \la \phi_m \,|\, \phi_m \ra_{A_m} = i \sigma_m \big( \phi_m, A_m \phi_m \big)
= i \sigma_m \big( A_m \phi_m, \phi_m \big) \:,
\eeq
where in the last step we took the complex conjugate and used that, according to~\eqref{sigma},
taking the complex conjugate of the symplectic form flips its arguments and gives a minus sign.
Polarizing~\eqref{topolarize}, we conclude that
\[ \sigma_m \big( \phi_m, A_m \tilde{\phi}_m \big) = 
\sigma_m \big( A_m \phi_m, \tilde{\phi}_m \big) \qquad \text{for all~$\phi_m, \tilde{\phi}_m
\in \Cisc(\scrM) \cap \Sol_m(\scrM)$}\:. \]
As a consequence,
\[ \la A_m \phi_m | \tilde{\phi}_m \ra_{A_m} = i \sigma \big(A_m \phi_m, A_m\tilde{\phi}_m \big) 
= i \sigma \big( \phi_m, A_m^2 \tilde{\phi}_m \big) = \la \phi_m | A_m \tilde{\phi}_m \ra_{A_m} \:, \]
giving the desired symmetry of~$A_m$ on~$\H_{A_m}$.
Using this symmetry property, the inequality~\eqref{unibound} simply means that the operators~$A_m$
are uniformly bounded on~$\H_{A_m}$. In order to see this, we first note that
uniform boundedness on~$\H_{A_m}$ can be expressed by 
demanding that for all~$\phi \in \Cisco(\scrM \times I)$ and~$m \in I$,
\[ \big| \la \phi_m | A_m \phi_m \ra_{A_m} \big| \leq c\: \|\phi_m\|_{A_m}^2 \:. \]
Expressing the scalar products via~\eqref{Amdef} in terms of the symplectic form gives
precisely the inequality~\eqref{unibound}.

Next, on families of solutions~$\psi, \phi \in \Cisco(\scrM \times I)$
we introduce the auxiliary scalar product by integrating over the mass parameter,
\beq \label{spm}
\la \psi | \phi \ra_A := \int_I \la \psi_m | \phi_m \ra_{A_m} \: m\,dm \:.
\eeq
Forming the completion gives the
Hilbert space~$(\H_A, \la .|. \ra_A)$. This Hilbert space can also be understood as the
direct integral of the Hilbert space~$\H_{A_m}$,
\[ \H_A = \int_{\oplus I} \H_{A_m}\: m\,dm \:. \]

Now the methods in~\cite{infinite} apply, giving the following results.
\begin{Thm} \label{thmsMOP}
The following statements are equivalent:
\bitem
\item[{\rm{(i)}}] There is
a constant~$c>0$ such that for all~$\phi, \tilde{\phi} \in \Cisco(\scrM \times I)$,
\[ 
\big| \la \p \phi | \p \tilde{\phi} \ra_{L^2(\scrM)} \big| \leq
c \int_I \, \| \phi_m\|_{A_m}\, \| \tilde{\phi}_m\|_{A_m}\: m\,dm \:. \]
\item[{\rm{(ii)}}] There is a constant~$c>0$ such that for all~$\phi, \tilde{\phi} \in \Cisco(\scrM \times I)$,
the following two relations hold:
\begin{align}
\big| \la \p \phi | \p \tilde{\phi} \ra_{L^2(\scrM)} \big| &\leq c\, \|\phi\|_A\, \|\tilde{\phi}\|_A \label{mb1} \\
\la \p T \phi | \p \tilde{\phi} \ra_{L^2(\scrM)} &= \la \p \phi | \p T \tilde{\phi} \ra_{L^2(\scrM)}\:. \label{mb2}
\end{align}
\item[{\rm{(iii)}}] There is a family of linear operators~$Q_m \in \Lin(\H_m)$ which are uniformly bounded,
\beq  \sup_{m \in I} \|Q_m\|_{\Lin(\H_m)} < \infty\:, \eeq \label{Qmdef2}
such that
\beq \label{Qmdef}
\la \p \phi | \p \tilde{\phi} \ra_{L^2(\scrM)} = \int_I \la \phi_m \,|\, Q_m \,\tilde{\phi}_m\ra_{A_m}\: m\,dm \qquad
\forall\: \phi, \tilde{\phi} \in \Cisco(\scrM \times I)\:.
\eeq
\eitem
\end{Thm}
\Proof Follows exactly as the proof of~\cite[Theorem~4.2]{infinite}.
\QED

The following proposition is proved exactly as~\cite[Proposition~4.2]{infinite}.
\begin{Prp} {\bf{(uniqueness of~$Q_m$)}} \label{prpunique}
The family~$(Q_m)_{m \in I}$ in the statement of Theorem~\ref{thmsMOP}
can be chosen such that for all~$\phi, \tilde{\phi} \in \Cisco(\scrM \times I)$, the expectation
value~$ f_{\phi, \tilde{\phi}}(m) := \la \phi_m | Q_m \tilde{\phi}_m\ra_{A_m}$ is continuous in~$m$,
\beq \label{flip}
f_{\phi, \tilde{\phi}} \in C^0_0(I) \:.
\eeq
The family~$(Q_m)_{m \in I}$ with the properties~\eqref{Qmdef} and~\eqref{flip} is unique.
Moreover, choosing two intervals~$\check{I}$ and~$I$ with~$m \in \check{I} \subset I$
and~$0 \not \in \overline{I}$, 
and denoting all the objects constructed in~$\check{I}$ with an additional check,
we have
\beq \label{ScS}
\check{Q}_m = Q_m \:.
\eeq
\end{Prp}

After these preparations, we are ready to represent the spacetime scalar product
in terms of the symplectic form.
\begin{Def}\label{defbmop} $\quad$ The Klein-Gordon operator~$P_m$ on the globally hyperbolic ma\-ni\-fold~$(\scrM, g)$
has the {\bf{bosonic mass oscillation property}} with respect to
the auxiliary operators~$(A_m)_{m \in I}$ if the following conditions hold:
\bitem
\item[{\rm{(i)}}] The operator~$T$ is symmetric on~$\H$, i.e.\
\beq \label{Tsymm}
\la T \phi | \tilde{\phi} \ra_\H = \la \phi | T \tilde{\phi} \ra_\H \qquad \text{for all$~\phi, \tilde{\phi} \in \Cisco(\scrM)$}\:.
\eeq
\item[{\rm{(ii)}}] The norms on~$\H$ and~$\H_A$ are equivalent, i.e.\ there is a constant~$c>0$ such that
\beq \label{normequiv}
\frac{1}{c}\: \|\phi\|_A \leq \|\phi\|_\H \leq c\: \|\phi\|_A \qquad \text{for all$~\phi \in \Cisco(\scrM)$}\:.
\eeq
\eitem
\end{Def}

\begin{Thm} Assume that the bosonic mass oscillation property holds with respect to
the auxiliary operators~$(A_m)_{m \in I}$. Then there is a unique family of
operators~$(\Sig_m)_{m \in I}$ on~$\H_{A_m}$ with the following properties:
\bitem
\item[{\rm{(i)}}] The spacetime scalar product is related to the symplectic form by
\beq \label{massdecomp}
\la \phi | \tilde{\phi} \ra_\H = i \int_I \sigma_m\big( \phi_m, \Sig_m \,\tilde{\phi}_m\big)\: m\,dm \qquad
\forall\: \phi, \tilde{\phi} \in \Cisco(\scrM \times I)\:.
\eeq
\item[{\rm{(ii)}}] For any~$\phi, \tilde{\phi} \in \Cisco(\scrM \times I)$, the
expectation value~$\sigma_m\big( \phi_m, \Sig_m \,\tilde{\phi}_m\big)$ is continuous in~$m$.
\item[{\rm{(iii)}}] The  family of linear operators~$\Sig_m \in \Lin(\H_m)$   are uniformly bounded,
\beq \label{Qmdef3}
\sup_{m \in I} \|\Sig_m\|_{\Lin(\H_m)} < \infty\:.
\eeq 
\eitem
Moreover, the operator~$\Sig_m$ does not depend on the choice of~$I$, in the sense that
choosing two intervals~$\check{I}$ and~$I$ with~$m \in \check{I} \subset I$,
then~$\Sig_m = \check{\Sig}_m$ for all~$m \in \check{I}$.
\end{Thm}
\Proof Using Cauchy-Schwarz,
the second inequality in~\eqref{normequiv} yields~\eqref{mb1}.
Moreover, \eqref{Tsymm} gives~\eqref{mb2}.
Therefore, Theorem~\ref{thmsMOP} applies, giving the mass decomposition~\eqref{Qmdef}.
Combining this formula with~\eqref{Amdef} yields
\[ \la \phi | \tilde{\phi} \ra_\H = i \int_I \sigma_m \big( \phi_m, A_m Q_m \,\tilde{\phi}_m\big) \:m\, dm\:. \]
Therefore, we obtain~\eqref{massdecomp} with~$\Sig_m := A_m Q_m\in\Lin(\H_m)$.
The operators~$\Sig_m$ are uniformly bounded in $m$ because of the
uniform boundedness of $A_m$  and $Q_m$ (see~\eqref{Qmdef2} and Definition~\ref{def31}~(ii)).
\QED

It remains to show that our results do not depend on the choice
of the auxiliary operators.
\begin{Thm} {\bf{(independence of the choice of~$A_m$)}}
Let~$(A_m)_{m \in I}$ and~$(A'_m)_{m \in I}$ be two families of auxiliary operators.
Then the corresponding Hilbert space
norms are equivalent, i.e.\ there is a constant~$C>0$ such that
\[ \frac{1}{C}\: \|\phi_m\|_{A'_m} \leq \|\phi_m\|_{A_m} \leq C\: \|\phi_m\|_{A'_m}
\qquad \text{for all~$\phi \in \Cisc(\scrM) \cap \Sol_m(\scrM)$}\:. \]
In other words, the Hilbert spaces~$\H_{A_m}$ and~$\H_{A'_m}$ are equivalent
as topological vector spaces. Moreover, the operators~$(\Sig_m)_{m \in I}$
corresponding to the choices of auxiliary operators~$(A_m)_{m \in I}$ and~$(A'_m)_{m \in I}$ coincide.
\end{Thm}
\Proof For clarity, we denote the operator~$\Sig_m$ corresponding to the
auxiliary operators~$(A_m)$ by~$\Sig_m^A$. Then, combining~\eqref{spm} with~\eqref{Amdef},
\[ \la \psi | \phi \ra_A := i \int_I  \sigma_m \big( \phi_m, A_m \tilde{\phi}_m \big) \: m\,dm \:, \]
and similarly for~$A'_m$. Using that, according to~\eqref{normequiv}, the corresponding norms are
both equivalent to the norm~$\|.\|_\H$, we conclude that there is a constant~$c$ such that
\[ \frac{1}{c} \int_I \|\phi_m\|^2_{A'_m} \: m\,dm \leq
\int_I \|\phi_m\|^2_{A_m} \: m\,dm \leq c \int_I  \|\phi_m\|^2_{A'_m} \: m\,dm \:. \]
Let~$\tilde{\phi}_m \in \Cisc(\scrM) \cap \Sol_m(\scrM)$. We extend~$\tilde{\phi}_m$ to a
family~$\tilde{\phi} \in \Cisco(\scrM\times I)$.
Next, we let~$\eta_\ell \in C^\infty_0(I)$ be a Dirac sequence which converges to the Dirac measure
at~$m$. Then, choosing~$\phi_m = \sqrt{\eta_\ell(m)}\: \tilde{\phi}_m$, we conclude that
\[ \frac{1}{c} \int_I \eta_\ell(m)\: \|\tilde{\phi}_m\|^2_{A'_m} \: m\,dm \leq
\int_I \eta_\ell(m)\: \|\tilde{\phi}_m\|^2_{A_m} \: m\,dm \leq c \int_I \eta_\ell(m)\: \|\tilde{\phi}_m\|^2_{A'_m} \: m\,dm \:. \]
From~\eqref{contin} we know that the functions~$\|\tilde{\phi}_m\|^2_{A_m}$ and~$\|\tilde{\phi}_m\|^2_{A'_m}$
are both continuous in~$m$. Therefore, we can take the limit~$\ell \rightarrow \infty$ to obtain
the pointwise inequality
\[ \frac{1}{c}\: \|\tilde{\phi}_m\|^2_{A'_m} \leq \|\tilde{\phi}_m\|^2_{A_m} \leq c\, \|\tilde{\phi}_m\|^2_{A'_m}
\qquad \text{for all~$\tilde{\phi}_m \in \Cisc(\scrM) \cap \Sol_m(\scrM)$}\:. \]
This establishes the desired equivalence of the norms. Hence we can identify~$\H_{A_m}$ and~$\H_{A'_m}$
as topological vector spaces.

In order to show uniqueness, we consider the
representations~\eqref{massdecomp} corresponding to~$A_m$ and~$A'_m$,
\[ \int_I \sigma_m\big( \phi_m, (\Sig^A_m - \Sig^{A'}_m)\,\phi_m\big)\: m\,dm = 0 \qquad
\text{for all~$\phi, \tilde{\phi}' \in \Cisco(\scrM \times I)$} \:.\]
Here the integrand is again continuous in~$m$, because for example
\[ \sigma_m\big( \phi_m, \Sig^A_m \phi_m\big)
= \sigma_m(\phi_m, A_m Q_m\,\phi_m\big) = \la \phi_m | Q_m \phi_m\ra_{A_m} \:, \]
which is continuous in view of~\eqref{flip}.
Therefore, choosing again~$\phi_m = \sqrt{\eta_\ell(m)}\: \tilde{\phi}_m$, we may take
the limit~$\ell \rightarrow \infty$ to obtain the pointwise identity
\[ \sigma_m\big( \tilde{\phi}_m, (\Sig^A_m - \Sig^{A'}_m)\,\tilde{\phi}_m\big) = 0
\qquad \text{for all~$\tilde{\phi}_m \in \Cisc(\scrM) \cap \Sol_m(\scrM)$}\:. \]
Polarizing and using a denseness argument, we conclude
that~$\Sig^A_m = \Sig^{A'}_m$ on~$\H_{A_m} \equiv \H_{A'
_m}$.
\QED

The integrand of~\eqref{massdecomp} gives a canonical scalar product on~$\Cisc(\scrM) \cap \Sol_m(\scrM)$,
\beq \label{sprodm}
\la \phi_m | \tilde{\phi}_m \ra_m := i \sigma_m\big( \phi_m, \Sig_m \,\tilde{\phi}_m\big) \:.
\eeq
Again dividing out the null space and taking the completion, we get the Hilbert spaces~$(\H_m, \la .|. \ra_m)$.
Taking their direct integrals gives back the Hilbert space~$\H$,
\[ \H = \int_{\oplus I} \H_m\: m\,dm \:. \]
It follows immediately from the above results that the norms of the
Hilbert spaces~$\H_m$ and~$\H_{A_m }$ are equivalent.

We finally remark that, if one can prove the existence or give an explicit expression for a family of
operators~$(\Sig_m)_{m \in I}$ satisfying~\eqref{massdecomp},
then the bosonic mass oscillation property given in Definition~\ref{defbmop} follows immediately
by choosing~$A_m = \Sig_m$.
 
\section{The Bosonic Signature Operator and Applications} \label{secsig}
The constructions of the previous section gave us for any~$m \in I$
a Hilbert space of solutions~$(\H_m, \la .|. \ra_m)$ with a canonical scalar product~\eqref{sprodm}.
We now choose~$m$ as the physical mass; from now on it will be kept fixed.
The operator~$\Sig_m$ appearing in the definition of this scalar product is
a bounded symmetric linear operator on~$\H_m$ which is uniquely defined by
the mass decomposition~\eqref{massdecomp}.

\begin{Def} The operator~$\Sig_m \in \Lin(\H_m)$ is
referred to as the {\bf{bosonic signature operator}}.
\end{Def}

\subsection{A Canonical Complex Structure} \label{seccomplex}
Being a bounded symmetric operator, we obtain the spectral decomposition
\[ \Sig_m = \int_{\sigma(\Sig_m)} \nu\: dE_\nu \]
with a projection-valued spectral measure~$dE$.
Assume that the operator~$\Sig_m$ has a trivial kernel,
\[ \ker \Sig_m = \{0\} \:. \]
Then  the operator~$J$ given by
\beq \label{Jdef}
J := i \,|\Sig_m|^{-1}\: \Sig_m = i \,\big( \chi_{(0, \infty)}(E) - \chi_{(-\infty,0)}(E) \big)
\eeq
defines a canonical complex structure on the solution space of the Klein-Gordon equation.
Indeed, the projection operators to its eigenspaces
\beq \label{projhol}
\chi^\hol := \frac{1}{2}\: (\1 - i J) \qquad \text{and} \qquad
\chi^\ah = \frac{1}{2}\: (\1 + i J) \:.
\eeq
map to subspaces of~$\H_m$ referred to as the {\em{holomorphic}} and {\em{anti-holomorphic}}
subspaces, respectively.

\subsection{The Bosonic Projector State} \label{secstate}
The canonical complex structure give rise to a distinguished quasi-free state,
as we now outline. Following the algebraic approach, we define the
{\em{algebra of fields}}~$\mathcal{A}(\scrM)$
as the free algebra generated by complex-valued test functions in~$\C_{0}(\scrM)$ 
divided by the ideal generated by the canonical commutation relations
\[ f \otimes g - g \otimes f = i G_m(f,g) \]
(here~$G$ is again the causal fundamental solution~\eqref{Gmform}).
A {\em{state}}~$\omega$ is a normalized and positive linear functional on the algebra of fields, i.e.
\beq \label{algstate}
\omega \::\:
\mathcal{A}(\scrM)\to\mathbb{C}\;\;\textrm{such that}\;\;\omega(\mathbb{\text{\rm{id}}})=1\;\;\textrm{and}\;\;\omega(a^*a)\geq 0\;\forall a\in\mathcal{A}(\scrM) \:.
\eeq
A state~$\omega$ is called {\em{quasi-free}} (or Gaussian) if the associated odd $n$-point functions all vanish,
while the even ones can be computed with the Wick rule
\[ \omega_{2n}(f_1 \dots f_m)=\sum\limits_{\sigma\in\mathcal{P}}\omega_2(f_{\sigma(1)}, f_{\sigma(2)}) \:\cdots\: \omega_2(f_{\sigma(2n-1)}, f_{\sigma(2n)}) \:, \]
where~$\mathcal{P}$ denotes all possible permutations of the set~$\{1,\dots,2n\}$ into a collection of elements~$\{\sigma(1),\dots,\sigma(2n)\}$ such that~$\sigma(2k-1)<\sigma(2k)$ for all~$k=1,\dots,n$.
According to this formula, a quasi-free state is uniquely determined by its two-point function~$\omega_2$.
For the {\em{bosonic projector state}} we choose
\beq \label{twopoint}
\omega_2(f,g) := i \sigma_m \big( G_m f, \chi^\hol\, G_m g \big) \:,
\eeq
where~$\chi^\hol$ is the projection to the holomorphic component~\eqref{projhol}.

\begin{Prp} The two-point function given in~\eqref{twopoint} defines a quasi-free bosonic state.
\end{Prp}
\Proof Our task is to verify the positivity statement in~\eqref{algstate} and the
compatibility with the canonical commutation relations. Before beginning, we form the
real Hilbert space~$(\H^\R_{A_m}, \la .|. \ra_m)$ formed of the real-valued solutions.
Restricting attention to these real-valued solutions, the scalar product and the symplectic form in~\eqref{sprodm}
are both real. Therefore, the operator~$i \Sig_m$ maps~$\H^\R_{A_m}$ to itself.
Using the spectral calculus, the same is true for the complex structure in~\eqref{Jdef},
\[ J : \H^\R_{A_m} \rightarrow \H^\R_{A_m} \:. \]
As a consequence, for real-valued $f$ and~$g$, we can decompose~\eqref{twopoint} into its real and
imaginary parts,
\begin{align*}
\omega_2(f,g) &= i \sigma_m \big( G_m f, \chi^\hol\, G_m g \big) 
= \frac{i}{2}\: \sigma_m \big( G_m f, (\1 - i J) G_m g \big) \\
&= \frac{i}{2}\: \sigma_m \big( G_m f, \Sig_m\: |\Sig_m|^{-1}\, G_m g \big) + \frac{i}{2}\: \sigma_m \big( G_m f, G_m g \big) \\
&= \frac{1}{2}\:\la G_m f\:|\: |\Sig_m|^{-1}\, G_m g \ra_m + \frac{i}{2}\: \sigma_m \big( G_m f, G_m g \big) \:.
\end{align*}
In particular, we conclude that the real part is positive semi-definite, and that the imaginary part satisfies the relation
\[ \im \omega_2(f,g) = \frac{1}{2}\:\sigma_m \big( G_m f, G_m g \big) \:. \]
Now we can apply Proposition~5.2.23~(b) in the
textbook~\cite{brunettibook} to obtain the result.
\QED

\section{The Limiting Case of Massless Fields} \label{secmassless}
In the constructions so far, it was essential that the scalar field was {\em{massive}}.
Indeed, we cannot expect that integrating over the mass generates the desired decay for large times
if we integrate over arbitrary small masses. This is the reason why we assumed that the interval~$I$
did not intersect a neighborhood of the origin. However, in many situations the bosonic signature operator~$\Sig_m$
can be defined for all~$m>0$. If this is the case, one can hope to extend the results to the massless case
by taking the limit~$m \searrow 0$.

In order to relate the solution spaces for different masses, we let~$\scrN$ be an arbitrary Cauchy
surface (we will show later that our results do not depend on the choice of this
Cauchy surface). We identify a solution~$\phi_m \in \Cisc(\scrM)$ with the corresponding Cauchy data
\[ \Phi := \begin{pmatrix} \phi_m \\[0.2em] i \partial_t \phi_m \end{pmatrix} \bigg|_\scrN
\in C^\infty_0(\scrN)^2 \:. \]
Clearly, the symplectic form~\eqref{sigma} can be expressed in terms of the Cauchy data;
we write for clarity~$\sigma_\scrN(\Phi, \tilde{\Phi})$. Moreover, the symplectic form is continuous
with respect to the Sobolev norm
\[ H^{1,2}(\scrN) \oplus L^2(\scrN) \:. \]
In what follows, it suffices to consider this norm locally on compact subsets of~$\scrN$.

\begin{Def} \label{definfrared}
The Klein-Gordon operator~$P_m$ is {\bf{infrared regular}} on the Cauchy surface~$\scrN$
if for every compact subset~$K \subset \scrN$, the restrictions of the bosonic signature operators
to~$K$ converge in norm, i.e.\ if the limit
\[ \lim_{m \searrow 0} \chi_K \Sig_m \big|_{H^{1,2}(K) \oplus L^2(K)} \qquad \text{exists in~$H^{1,2}(K) \oplus L^2(K)$}\:. \]
\end{Def}
Under the above assumptions, we can define the sesquilinear form
\beq \label{Hil0}
\la . | . \ra_0 \::\: C^\infty_0(\scrN)^2 \times C^\infty_0(\scrN)^2 \rightarrow \C \:,\qquad
\la \Phi | \tilde{\Phi} \ra_0 := \lim_{m \searrow 0} i \sigma_\scrN \big( \Phi \big| \Sig_m \tilde{\Phi} \big) \:.
\eeq
Being the limit of scalar products, this sesquilinear form is positive semi-definite. Dividing out the null space
and taking the completion, we obtain a Hilbert space, which we denote by~$(\H_0, \la .|. \ra_0)$.

\begin{Prp} If the Klein-Gordon operator~$P_m$ is infrared regular on the Cauchy surface~$\scrN$,
then it is also infrared regular on any other Cauchy surface. Moreover, the limit in~\eqref{Hil0}
does not depend on the choice of the Cauchy surface.
\end{Prp}
\Proof Let~$\scrN'$ be another Cauchy surface. We describe the time evolution operators from~$\scrN$
to~$\scrN'$ and from~$\scrN'$ to~$\scrN$ by
\[ U_m^{\scrN', \scrN} : C^\infty_0(\scrN)^2 \rightarrow C^\infty_0(\scrN')^2
\qquad \text{and} \qquad
U_m^{\scrN, \scrN'} : C^\infty_0(\scrN')^2 \rightarrow C^\infty_0(\scrN)^2 \:. \]
These mappings are symplectomorphisms. Moreover, standard energy estimates show that these
mappings are continuous in~$H^{1,2}_\loc \oplus L^2_\loc$ on the respective Cauchy surfaces
(for details see for example~\cite[Section~5.3]{john} or~\cite[Chapter~13]{intro}).

Suppose we consider the symplectic forms in~\eqref{Hil0} on the Cauchy surface~$\scrN'$.
Using that the time evolution operator preserves the symplectic form, we can also compute it on~$\scrN$,
\[ \sigma_{\scrN'} \big( \Phi \,\big|\, \Sig_m \tilde{\Phi} \big) = 
\sigma_{\scrN} \big( U_m^{\scrN, \scrN'} \Phi \,\big|\, ( U_m^{\scrN, \scrN'} \Sig_m 
U_m^{\scrN', \scrN} ) \,U_m^{\scrN, \scrN'}
\tilde{\Phi} \big) \:. \]
Due to causality, there is a compact set~$K \subset \scrN'$
(which clearly depends on~$\Phi$ and~$\tilde{\Phi}$) such that the supports of 
all the waves on the right hand side of the last equation are contained in~$K$ for all~$m$.
Therefore, we can take the limit~$m \searrow 0$ using that these vectors converge in the Hilbert space
in~$H^{1,2}(K) \oplus L^2(K)$, and that the operators~$\Sig_m$ converge in norm on this Hilbert space.
\QED

As a result of the above construction, we have two sesquilinear forms on
the massless solutions~$\Cisc(\scrM) \cap \Sol_0(\scrM)$: The symplectic form~$\sigma_0$
and the scalar product~$\la .|. \ra_0$. Since we took a rather weak limit, it is a-priori not clear
whether these sesquilinear forms are bounded relative to each other. If they are, one can
express one sesquilinear form relative to the other, giving rise to operators on the respective spaces.
The spectral decompositions of these operators again give rise to a splitting of the solution space
into distinguished subspaces. In order to illustrate how this can be done, assume that the
symplectic form is bounded pointwise by the scalar product, i.e.\ that for every~$\phi_0 \in
\Cisc(\scrM) \cap \Sol_0(\scrM)$ there is a constant~$c>0$ such that
\[ \big| \sigma_0(\phi_0, \tilde{\phi}_0) \big| \leq c\: \|\tilde{\phi}_0\|_0 \qquad \text{for
all~$\tilde{\phi}_0 \in \Cisc(\scrM) \cap \H_0$}\:. \]
Then the Fr{\'e}chet-Riesz theorem gives a symmetric linear operator
\[ \Sig_0^{-1} \::\: \Cisc(\scrM) \cap \H_0 \rightarrow \H_0 \]
which is uniquely defined by the property
\[ \sigma_0(\phi_0, \tilde{\phi}_0) = \big\la \phi_0 \,\big|\, \Sig_0^{-1}\,\tilde{\phi}_0 \big\ra_0 \qquad \text{for all~$\tilde{\phi}_0 \in \Cisc(\scrM) \cap \H_0$}\:. \]
Other variations of this constructions can be used depending on the specific properties of the 
resulting sesquilinear forms.

For clarity, we finally point out that the above constructions are {\em{not conformally invariant}}
in the sense that, even when applied to the conformal wave equation, the resulting scalar product~$\la .|. \ra_0$
will transform in an intricate (in general nonlocal) way under conformal transformations.
This can be understood from the fact that the mass oscillation property makes it necessary to consider
families of {\em{massive}} solutions, thereby breaking the conformal symmetry.

\section{Example: Minkowski Space}\label{secmink}
In this section, we verify by direct computation that the bosonic mass oscillation property holds in Minkowski space.
Moreover, we derive an explicit formula for the bosonic signature operator.
As we shall see, the resulting canonical complex structure will give us back the usual frequency
splitting. This is an important check showing that the bosonic signature operator gives physically
sensible results.

We consider Minkowski space, i.e.\ the spacetime~$\scrM=\R\times\R^3$ with the line element
\[ ds^2 =dt^2-(dx^1)^2 - (dx^2)^2 - (dx^3)^2 \:. \]
We let~$\phi_m$ be a solution of the Klein-Gordon equation for a given mass~$m>0$,
\[ (\square +m^2)\,\phi_m(x)=0 \:. \]
For initial data on~$\R^3$ at time~$t=0$ denoted by
\[ \phi_{m,0}(\mathbf{x}) := \phi_m(x)\big|_{(0,\mathbf{x})} \qquad \text{and} \qquad
\pi_{m,0}(\mathbf{x}) := i \partial_t \phi_m(x)\big|_{(0,\mathbf{x})} \:, \]
the Cauchy problem has a unique solution
(here~$\mathbf{x}=(x^1, x^2, x^3) \in\R^3$, and~$\pi_{m,0}(\mathbf{x})$ is the canonical momentum).

For our purposes, it is most convenient to write the solution of the Klein-Gordon equation as a Fourier
integral over the upper and lower mass shell,
\[ \phi_m(x)=\int \frac{d^4k}{(2 \pi)^4}\: \delta(k^2-m^2)\:\epsilon(k^0)\:\varphi_m(k)
\:e^{-ikx} \:. \]
Then initial data is computed by
\begin{align}
\phi_{m,0}(\mathbf{x}) &=\int \frac{d^4k}{(2 \pi)^4} \:\delta(k^2-m^2)\:\epsilon(k^0)\:\varphi_m(k)\: e^{i\mathbf{k}\mathbf{x}} \notag \\
&=\int \frac{d^3k}{(2 \pi)^3} \: \frac{1}{4 \pi\,\omega_{{k}}} \:\big( \varphi_m( \omega_k,\mathbf{k})-\varphi_m( -\omega_k, \mathbf{k}) \big) \label{phim0} \\
\pi_{m,0}(\mathbf{x}) 
&=\int \frac{d^3k}{(2 \pi)^3} \: \frac{1}{4 \pi} \:\big( \varphi_m( \omega_k,\mathbf{k}) + \varphi_m( -\omega_k, \mathbf{k}) \big) \:, \label{pim0}
\end{align}
where~$\omega_k := \sqrt{\mathbf{k}^2+m^2}$.
Next, the symplectic form~\eqref{sigmaintro} takes the form
\[ \sigma_m(\phi_m , \tilde{\phi}_m)
= i \int \Big(
\overline{\pi_{m,0}(\mathbf{x})}\, \tilde{\phi}_{m,0}(\mathbf{x})+\overline{\phi_{m,0}(\mathbf{x})}\, \tilde{\pi}_{m,0}(\mathbf{x})
\Big)\:d^3x \:. \]
Applying Plancherel's theorem in the spatial variables, we obtain
\begin{align}
&\sigma_m(\phi_m, \tilde{\phi}_m) \notag \\
&=  i \int \frac{d^3k}{(2 \pi)^3}\: \frac{1}{16 \pi^2\, \omega_k}\Big(
\overline{\big( \varphi_m( \omega_k,\mathbf{k}) + \varphi_m( -\omega_k, \mathbf{k}) \big)}
\big( \tilde{\varphi}_m( \omega_k,\mathbf{k})- \tilde{\varphi}_m( -\omega_k, \mathbf{k}) \big) \notag \\
&\qquad\qquad\qquad\qquad\;\;+ \overline{\big( \varphi_m( \omega_k,\mathbf{k}) - \varphi_m( -\omega_k,\mathbf{k}) \big)}
\big( \tilde{\varphi}_m( \omega_k,\mathbf{k}) + \tilde{\varphi}_m( -\omega_k, \mathbf{k}) \big) \Big) \notag \\
&= \int \frac{d^3k}{(2 \pi)^3}\: \frac{i}{8 \pi^2\, \omega_k}
\Big( \overline{\varphi_m( \omega_k, \mathbf{k})}\, \tilde{\varphi}_m( \omega_k, \mathbf{k})
- \overline{\varphi_m(-\omega_k, \mathbf{k})}\, \tilde{\varphi}_m(-\omega_k, \mathbf{k}) \Big) \:. \label{sympultra}
\end{align}

We next consider a family of solutions~$\phi = (\phi_m)_{m \in I} \in \Cisco(\scrM \times I)$.
Integrating over the mass according to~\eqref{pdef} gives
\begin{align*}
(\p \psi)(x) &= \int_I m\, dm \int \frac{d^4k}{(2 \pi)^4}\: \delta(k^2-m^2)\: \epsilon(k^0) \:\varphi_m(k)\: e^{-ikx} \\
&= \frac{1}{2} \int \frac{d^4k}{(2 \pi)^4}\: \epsilon(k^0) \:\varphi_m(k) \Big|_{m=\sqrt{k^2}}\;e^{-ikx}
\end{align*}
(note that the assumption~$\phi = (\phi_m)_{m \in I} \in \Cisco(\scrM \times I)$ implies that
the function~$\varphi_m(k)$ vanishes unless~$\sqrt{k^2} \in I$).
Applying Plancherel's theorem in spacetime, it follows that
\begin{align*} 
&\langle \p \phi,\p \tilde{\phi} \rangle_{L^2(\R^4)} =\frac{1}{4} \int \frac{d^4k}{(2 \pi)^4}\:  \overline{\varphi_m(k)}\,\tilde{\varphi}_m(k)
\Big|_{m=\sqrt{k^2}} \\
&=\frac{1}{2} \int_I m \, dm \int \frac{d^4k}{(2 \pi)^4}\: \delta(k^2-m^2)\: \overline{\varphi_m(k)}\, \tilde{\varphi}_m(k)
\Big|_{m=\sqrt{k^2}} \\
&= \int_I\,m\, dm \,\int \frac{d^3\mathbf{k}}{(2 \pi)^3}\: \frac{1}{8 \pi\, \omega_k}\: 
\Big( \overline{\varphi_m( \omega_k, \mathbf{k})}\, \tilde{\varphi}_m( \omega_k, \mathbf{k})
+ \overline{\varphi_m(-\omega_k, \mathbf{k})}\, \tilde{\varphi}_m(-\omega_k, \mathbf{k})  \Big) \:.
\end{align*}
Comparing with~\eqref{sympultra}, we conclude that
\[ \langle \p \phi,\p \tilde{\phi} \rangle_{L^2(\R^4)} = 
i \int_I \sigma_m\big( \phi_m, \Sig_m \,\tilde{\phi}_m\big)\: m\,dm \:, \]
where the operator~$\Sig_m$ acts on the solutions of positive and negative frequency by
\beq \label{Sultra}
\Sig_m \,\varphi_m( \pm \omega_k, \mathbf{k}) =  \mp \pi\: \varphi_m( \pm \omega_k, \mathbf{k}) \:.
\eeq

We finally rewrite the bosonic signature operator as an operator acting on the initial data.
To this end, from~\eqref{phim0} and~\eqref{pim0} we read off that
\begin{align*}
\begin{pmatrix} \hat{\phi}_{m,0}(\mathbf{k}) \\ \hat{\pi}_{m,0}(\mathbf{k}) \end{pmatrix}
&= \frac{1}{4 \pi} \begin{pmatrix} 1/\omega_k & -1/\omega_k \\ 
1 & 1 \end{pmatrix}
\begin{pmatrix} \varphi_m( \omega_k,\mathbf{k}) \\ \varphi_m(-\omega_k,\mathbf{k}) \end{pmatrix} \\
\begin{pmatrix} \varphi_m( \omega_k,\mathbf{k}) \\ \varphi_m(-\omega_k,\mathbf{k}) \end{pmatrix}
&= 2 \pi \begin{pmatrix} \omega_k & 1 \\ 
-\omega_k & 1 \end{pmatrix} \begin{pmatrix} \hat{\phi}_{m,0}(\mathbf{k}) \\ \hat{\pi}_{m,0}(\mathbf{k})
\end{pmatrix} ,
\end{align*}
where~$\hat{\phi}_{m,0}$ and~$\hat{\pi}_{m,0}$ denote the spatial Fourier transforms.
Thus, using~\eqref{Sultra},
\begin{align*}
\Sig_m \begin{pmatrix} \hat{\phi}_{m,0}(\mathbf{k}) \\ \hat{\pi}_{m,0}(\mathbf{k}) \end{pmatrix}
&= \frac{1}{4\pi} \begin{pmatrix} 1/\omega_k & -1/\omega_k \\ 
1 & 1 \end{pmatrix}
\begin{pmatrix} -\pi \varphi_m( \omega_k,\mathbf{k}) \\ \pi \varphi_m(-\omega_k,\mathbf{k}) \end{pmatrix} \\
&= \frac{\pi}{2} \begin{pmatrix} 1/\omega_k & -1/\omega_k \\ 
1 & 1 \end{pmatrix}
\begin{pmatrix} -\omega_k & -1 \\ 
-\omega_k & 1 \end{pmatrix} \begin{pmatrix} \hat{\phi}_{m,0}(\mathbf{k}) \\ \hat{\pi}_{m,0}(\mathbf{k})
\end{pmatrix} \:.
\end{align*}
We thus obtain the simple formula
\beq \label{Sminkfinal}
\Sig_m \begin{pmatrix} \hat{\phi}_{m,0}(\mathbf{k}) \\ \hat{\pi}_{m,0}(\mathbf{k}) \end{pmatrix}
= -\pi \begin{pmatrix} 0 & 1/\omega_k \\ 
\omega_k & 0 \end{pmatrix} \begin{pmatrix} \hat{\phi}_{m,0}(\mathbf{k}) \\ \hat{\pi}_{m,0}(\mathbf{k}) \end{pmatrix} \:.
\eeq

The scalar product~\eqref{sprodm} can be computed most easily by using~\eqref{Sultra}
in~\eqref{sympultra}. We thus obtain
\beq \label{sprodmMink}
\la \phi_m | \tilde{\phi}_m \ra_m = \int \frac{d^3k}{(2 \pi)^3}\: \frac{1}{8 \pi\, \omega_k}
\Big( \overline{\varphi_m( \omega_k, \mathbf{k})}\, \tilde{\varphi}_m( \omega_k, \mathbf{k})
+ \overline{\varphi_m(-\omega_k, \mathbf{k})}\, \tilde{\varphi}_m(-\omega_k, \mathbf{k}) \Big) \:.
\eeq
This is positive definite and Lorentz invariant, as desired.
Moreover, one sees immediately that in the limit~$m \searrow 0$, this scalar product
as well as the bosonic signature operators converge to
\begin{gather*}
\Sig_0 \,\varphi_0( \pm |\mathbf{k}|, \mathbf{k}) =  \mp \pi\: \varphi_m( \pm |\mathbf{k}|, \mathbf{k}) \\
\la \phi_0 | \tilde{\phi}_0 \ra_0 = \int \frac{d^3k}{(2 \pi)^3}\: \frac{1}{8 \pi\, |\mathbf{k}|}
\Big( \overline{\varphi_m( |\mathbf{k}|, \mathbf{k})}\, \tilde{\varphi}_m( |\mathbf{k}|, \mathbf{k})
+ \overline{\varphi_m(-|\mathbf{k}|, \mathbf{k})}\, \tilde{\varphi}_m(-|\mathbf{k}|, \mathbf{k}) \Big) \:.
\end{gather*}
Using the notion of Definition~\ref{definfrared}, the Klein-Gordon equation in Minkowski space
is infrared regular. Moreover, the operator~$\Sig_0$ is again bounded on~$(\H_0, \la .|. \ra_0)$.
Therefore, all our results apply as well to the scalar wave equation in Minkowski space.

\section{Example: Ultrastatic Spacetimes} \label{secultra}
In this section we prove that the bosonic mass oscillation property holds for 
the Klein-Gordon operator in ultrastatic spacetimes.
We also compute the bosonic signature operator explicitly and verify that
the resulting canonical complex structure reproduces the usual frequency splitting.
For simplicity, we only consider the case~$\xi=0$ without coupling to scalar curvature.
Thus we let~$(\scrM,g)$ be a $k$-dimensional globally hyperbolic spacetime
which is ultrastatic in the sense that
it is the product~$\scrM= \R \times \scrN$ with a metric of the form
\[ ds^2 = dt^2 -g_\scrN \:, \]
where $g_\scrN$ is a Riemannian metric on~$\scrN$.
The global hyperbolicity of~$(\scrM, g)$ implies that~$(\scrN, g_\scrN)$ is complete.
In preparation, we formulate the Klein-Gordon equation in the {\em{Hamiltonian form}},
following the procedure in~\cite{dimock-kay}.
We denote spacetime points~$x \in \scrM$ by~$x=(t,\x)$ with~$\x \in \scrN$.
We introduce the two-component vector~$\Phi_m$ by
\[ \Phi_m(t,\x) = \begin{pmatrix} \phi_m(t,\x) \\ i \partial_t \phi_m(t,\x) \end{pmatrix} \:. \]
Then the Klein-Gordon equation can be written as
\[ i \partial_t \Phi_m = H_m \Phi_m \]
with the {\em{Hamiltonian}}~$H_m$ given by
\[ H_m = \begin{pmatrix} 0 & 1 \\ -\Delta_\scrN + m^2 & 0 \end{pmatrix} \:. \]
Moreover, the symplectic form~\eqref{sigma} takes the form
\begin{align*}
\sigma_\scrN \big( \Phi_m, \tilde{\Phi}_m \big) &= \int_\scrN \Big( 
\overline{\partial_t \phi_m}\: \tilde{\phi}_m - \overline{\phi_m}\: \partial_t \tilde{\phi}_m \Big) \:d\mu_\scrN(x) \\
&= i \int_\scrN \Big\la \Phi_m , \begin{pmatrix} 0 & 1 \\ 1 & 0 \end{pmatrix} \tilde{\Phi}_m \Big\ra_{\C^2}\:d\mu_\scrN(x) \:.
\end{align*}

In what follows, we shall work with the spectral calculus for the operator
\[ K_m^2 := -\Delta_\scrN + m^2 \]
with domain~$C^\infty_0(\scrN)$ on the Hilbert space~$L^2(\scrN)= L^2(\scrN, d\mu_\scrN)$.
Using Chernoff's method~\cite{chernoff73}, one sees that the operator~$K_m^2$ is essentially self-adjoint
(for details see~\cite[Theorem~7.1]{kay} or~\cite{much+oeckl}).
We denote the self-adjoint extension again by~$K_m^2$. Taking the square root,
we obtain the spectral decomposition
\[ K_m = \int_{\sigma(K_m)} k\: dE_k \:, \]
where~$E_k$ is a projection-valued spectral measure.
Moreover, we know that
\[ \sigma(K_m) = \R^+ \:. \]
Consequently, the Hamiltonian has the spectral decomposition
\[ H_m = \begin{pmatrix} 0 & 1 \\ K_m^2+m^2 & 0 \end{pmatrix} 
= \int_0^\infty \begin{pmatrix} 0 & 1 \\ k^2+m^2 & 0 \end{pmatrix} \: dE_k \:. \]
By direct computation, one verifies that
\[ e^{-i t H_m} = \int_0^\infty \begin{pmatrix} \cos(\omega_k t)
& \displaystyle -i \:\frac{\sin(\omega_k t)}{\omega_k} \\[1em]
-i \omega_k\: \sin(\omega_k t) & \cos(\omega_k t) \end{pmatrix} \: dE_k \:, \]
where we introduced the notation
\[ \omega_k := \sqrt{k^2+m^2} \:. \]
Hence, given initial data~$\Phi_{m,0} \in C^\infty_0(\scrN)^2$, the solution of the Cauchy problem takes the form
\begin{align*}
\Phi_m(t) &= \int_0^\infty \begin{pmatrix} \cos(\omega_k t)
& \displaystyle -i \:\frac{\sin(\omega_k t)}{\omega_k} \\[0.7em]
-i \omega_k\: \sin(\omega_k t) & \cos(\omega_k t) \end{pmatrix} \: dE_k\: \Phi_{m,0} \\
&= \frac{1}{2} \int_0^\infty \bigg( e^{-i \omega_k t} \:
\begin{pmatrix} 1 & 1/\omega_k \\
\omega_k & 1 \end{pmatrix} 
+ e^{i \omega_k t} \begin{pmatrix} 1
& \displaystyle -1/\omega_k \\
-\omega_k & 1 \end{pmatrix}  \bigg) \: dE_k\: \Phi_{m,0} \:.
\end{align*}
In order to simplify the following computation, we restrict attention to the case
of {\em{non-compact Cauchy surface}} (the case of compact Cauchy surfaces can be treated
similarly by replacing the $\omega$-integrals by sums).
The   spectral measure is absolutely continuous
with respect to the Lebesgue measure, i.e.\ $dE_k = E_k \,dk$
(for details see~\cite{xavier}). 
Introducing~$\omega$ as an
integration variable, we obtain
\[ \Phi_m(t)  = 
\frac{1}{2} \int_m^\infty \bigg( e^{-i \omega t} \:
\begin{pmatrix} 1 & 1/\omega \\
\omega & 1 \end{pmatrix} 
+ e^{i \omega t} \begin{pmatrix} 1
& \displaystyle -1/\omega \\
-\omega & 1 \end{pmatrix}  \bigg) \: E_{k_\omega}\: \Phi_{m,0}\:\frac{\omega}{k_\omega}\: d\omega \:, \]
where~$k_\omega:= \sqrt{\omega^2-m^2}$. Therefore, the Fourier transform in~$t$ is computed by
\begin{align*}
\Phi_m(t) &= \int_{-\infty}^\infty \frac{d\omega}{(2 \pi)}\: \hat{\Phi}_m(\omega)\: e^{-i \omega t} \qquad \text{with} \\
\hat{\Phi}_m(\omega) &= \pi\: \frac{|\omega|}{k_\omega}\: \begin{pmatrix} 1 & 1/\omega \\
\omega & 1 \end{pmatrix} E_{k_\omega}\: \Phi_{m,0} \:.
\end{align*}
Hence
\begin{align*}
\big( \p \hat{\Phi} \big)(\omega)
&= \pi \int_{I \cap [0, |\omega|]} \frac{|\omega|}{\sqrt{\omega^2-m^2}}\: \begin{pmatrix} 1 & 1/\omega \\
\omega & 1 \end{pmatrix} E_{\sqrt{\omega^2-m^2}}\: \Phi_{m,0}\: m\,dm \\
&= \pi \int_0^\omega \chi_I(m_k) \:|\omega|\: \begin{pmatrix} 1 & 1/\omega \\
\omega & 1 \end{pmatrix} dE_k\: \Phi_{m_k,0} \:,
\end{align*}
where we transformed back to a $k$-integral and set~$m_k= \sqrt{\omega^2-k^2}$. Now we can apply Plancherel's theorem in the time variable to obtain
\begin{align*}
&\la \p \phi \,|\, \p \tilde{\phi} \ra_{L^2(\scrM, d\mu_\scrM)}
= \int_{-\infty}^\infty \la \p \phi(t) \,|\, \p \tilde{\phi}(t) \ra_{L^2(\scrN)}\: dt
= \int_{-\infty}^\infty \frac{d\omega}{2 \pi}\: \la \p \hat{\phi}(\omega) \,|\, \p \hat{\tilde{\phi}}(\omega) \ra_{L^2(\scrN)} \\
&= \frac{\pi}{2} \int_{-\infty}^\infty d\omega
\int_0^\omega \chi_I(m_k) \:\omega^2\,
\Big\la \big(\Phi_{m_k,0}^1 + \frac{1}{\omega}\: \Phi_{m_k,0}^2 \big)\,\Big|\, dE_k 
\big(\tilde{\Phi}_{m_k,0}^1 + \frac{1}{\omega}\: \tilde{\Phi}_{m_k,0}^2 \big) \Big\ra_{L^2(\scrN)} \\
&= \pi \int_{0}^\infty d\omega
\int_0^\omega \chi_I(m_k) \:\Big( \omega^2
\big\la \Phi_{m_k,0}^1 \big|\, dE_k \tilde{\Phi}_{m_k,0}^1 \big\ra_{L^2(\scrN)} 
+ \big\la \Phi_{m_k,0}^2 \big|\, dE_k \tilde{\Phi}_{m_k,0}^2 \big\ra_{L^2(\scrN)} \Big) \\
&= \pi \int_{0}^\infty d\omega
\int_0^\omega \chi_I(m_k) \:
\big\la \Phi_{m_k,0} \big|\, \begin{pmatrix} \omega^2 & 0 \\ 0 & 1 \end{pmatrix} dE_k \tilde{\Phi}_{m_k,0}
\big\ra_{L^2(\scrN)} \\
&= \pi \int_I dm
\int_0^\infty \frac{m}{\omega_k}\:
\big\la \Phi_{m,0} \big|\, \begin{pmatrix} \omega_k^2 & 0 \\ 0 & 1 \end{pmatrix} dE_k \tilde{\Phi}_{m,0}
\big\ra_{L^2(\scrN)} \\
&= \pi \int_I m\, dm \int_0^\infty 
\big\la \Phi_{m,0} \big|\, \begin{pmatrix} 0 & 1 \\ 1 & 0 \end{pmatrix}
\begin{pmatrix} 0 & 1/\omega_k \\ \omega_k & 0 \end{pmatrix} dE_k \tilde{\Phi}_{m,0}
\big\ra_{L^2(\scrN)} \\
&=-i\pi \int_I m\, dm \int_0^\infty 
\sigma_m \Big( \Phi_{m,0} \:,
\begin{pmatrix} 0 & 1/\omega_k \\ \omega_k & 0 \end{pmatrix} dE_k \tilde{\Phi}_{m,0}
\Big) \\
&=i \int_I \sigma_m \big( \Phi_{m,0}, \Sig_m\, \tilde{\Phi}_{m,0} \big)\: m\,dm
\end{align*}
with
\beq \label{Sigultra}
\Sig_m\, \Phi_{m,0} := -\pi \int_0^\infty \begin{pmatrix} 0 & 1/\omega_k \\ \omega_k & 0 \end{pmatrix} dE_k\:
\Phi_{m,0} \:.
\eeq
In the case of Minkowski space, this formula gives back~\eqref{Sminkfinal}.
The eigenspaces of this operator are precisely the solutions of positive and negative energy,
respectively.

Now the scalar product~\eqref{sprodm} takes the form
\[ \la \phi_m \,|\, \tilde{\phi}_m \ra_m 
= \pi \int_0^\infty \frac{1}{\omega_k}\:
\big\la \Phi_{m,0} \big|\, \begin{pmatrix} \omega_k^2 & 0 \\ 0 & 1 \end{pmatrix} dE_k \tilde{\Phi}_{m,0}
\big\ra_{L^2(\scrN)} \:. \]
This is positive definite, as desired. Similar as described at the end of the previous section
in Minkowski space, in an ultrastatic spacetime the Klein-Gordon equation is again
infrared regular. More precisely, in the limit~$m \searrow 0$ of the scalar product and the bosonic signature
operator converge to
\begin{align*}
\la \phi_0 \,|\, \tilde{\phi}_0 \ra_0 
&= \pi \int_0^\infty \frac{1}{\omega_k}\:
\big\la \Phi_{m,0} \big|\, \begin{pmatrix} k^2 & 0 \\ 0 & 1 \end{pmatrix} dE_k \tilde{\Phi}_{m,0}
\big\ra_{L^2(\scrN)} \\
\Sig_0 \Phi_{0,0} &= -\pi \int_0^\infty \begin{pmatrix} 0 & 1/k \\ k & 0 \end{pmatrix} dE_k \:\Phi_{0,0}\:. 
\end{align*}
In this way, in ultrastatic spacetimes all our results apply to the scalar wave equation as well.

\section{Discussion and Outlook}
In the present paper we gave a general construction of the bosonic signature operator in
globally hyperbolic spacetimes. For the case of ultrastatic spacetimes, the
complex structure defined by the bosonic signature operator in~\eqref{Jdef}, i.e.\ $$J := i \,|\Sig_m|^{-1}\,\Sig_m $$
agrees with the usual frequency splitting as obtained in in~\cite{kay} and~\cite{much+oeckl2}.
As a consequence, the resulting bosonic projector state is a Hadamard state.
The main advantage of working with the bosonic signature operator is that the construction
also applies to general time-dependent situations, in which the time evolution is not unitarily implementable
Hamiltonian (see \cite{helfer}). The only condition needed is that the bosonic mass oscillation property
holds (see Definition~\ref{defbmop}). In this way, we obtain a canonical quantum state
which does not depend on an observer. Instead, this state is determined by the global geometry
of spacetime. We remark that, in the setting of a scattering process, the bosonic signature state
does not coincide with the vacuum of the observer in the past or future, but instead it can be considered as being
an ``interpolation'' between the scattering states
(for a discussion of this point in the fermionic context see~\cite[Section~5]{sea}).

In the time-dependent situation, it is unknown whether the bosonic projector state is a Hadamard state.
At present, there is no general argument to prove the Hadamard property (in particular, due to the
dependence on the global geometry, it does not seem possible to apply gluing arguments
as in~\cite{fulling+sweeny+wald, fulling+narcowich+wald}).
Also, at present there is no general method for verifying if a given spacetime has the 
bosonic mass oscillation property by constructing a suitable family of
auxiliary operators~$(A_m)_{m \in I}$.
Instead, both the bosonic mass oscillation
property and the Hadamard property must be verified for different classes of spacetimes in a case to case study.
We expect that, similar as proven in the fermionic case
in~\cite{hadamard, planewave, desitter, sigbh}, the bosonic projector state should be Hadamard
in many physically interesting situations. These questions are part of an ongoing research program.
Another direction of research is to extend our construction to spacetimes with bifurcate Killing horizons,
as studied in the fermionic case in~\cite{rindler, sigbh, drago+murro}.

\Thanks{{{\em{Acknowledgments:}} We are grateful for support by the German Science Foundation (DFG)
within the Priority Program SPP 2026 ``Geometry at Infinity.''
We would like to thank the referees for valuable comments.

\providecommand{\bysame}{\leavevmode\hbox to3em{\hrulefill}\thinspace}
\providecommand{\MR}{\relax\ifhmode\unskip\space\fi MR }
\providecommand{\MRhref}[2]{%
  \href{http://www.ams.org/mathscinet-getitem?mr=#1}{#2}
}
\providecommand{\href}[2]{#2}

\end{document}